## Forward and Inverse Kinematics Seamless Matching Using Jacobian

Z. BHATTI++, A. SHAH, F. SHAHIDI, M. KARBASI

Khulliyyah of Information and Communication Technology, International Islamic University Malaysia
asadullah@iium.edu.my, shfarruh@kict.iium.edu.my, mostafa.karbasi@live.iium.edu.my

**Abstract:** In this paper the problem of matching Forward Kinematics (FK) motion of a 3 Dimensional (3D) joint chain to the Inverse Kinematics (IK) movement and vice versa has been addressed. The problem lies at the heart of animating a 3D character having controller and manipulator based rig for animation within any 3D modeling and animation software. The seamless matching has been achieved through the use of pseudo-inverse of Jacobian Matrix. The Jacobian Matrix is used to determine the rotation values of each joint of character body part such as arms, between the inverse kinematics and forward kinematics motion. Then moving the corresponding kinematic joint system to the desired place, automatically eliminating the jumping or popping effect which would reduce the complexity of the system.

## 1. INTRODUCTION

In the field of motion dynamics, there are generally two methods used, Kinematics and Kinetics. Kinematics is the study of motion, describing how a hierarchical skeletal structure move irrespective of the real world forces that may or may not have an effective impact on the motion gaits. Whereas, the study involving the effect of real world forces and torques on the dynamic motion characteristics of a structure is called Kinetics. In the field of computer animation both principles are used extensively to produce realistic dynamic motions of characters. However in this paper the focus is on the use of kinematics in motion dynamics.

The research on kinematics and kinetics is done in various fields from Medical science (Apriantono, *et al.,* 2006, Roepstorff, *et al.,* 1999 and Jonathan, *et al.,* 2008), biomechanics (Robin, *et al.,* 2006 and Bruce, *et al.,* 2003), aviation (Stirling, *et al.,* 2009), robotics (Kulpa, *et al.,* 2005, kuffner, *et al.,* 2005, Tejomurtula, *et al.,* 1999, Xia, *et al.,* 2005 and Li, *et al.,* 2001) and computer animation (Grochow, *et al.,* 2004 and Rose III, *et al.,* 2001). The common use of kinematics in the field of computer animation is for manipulating character's limbs which are defined through hierarchal tree of rotational joints connected with each other through rigid links (Parent, 2008). Two types of kinematics based approaches are used in positioning each hierarchy or rotational joint limbs, one is known as Forward Kinematics (FK) and the other is Inverse Kinematics (IK).

### 1.1 Forward Kinematics

In FK system the animator must specify all the parameters for degree of rotation and their order for each joint in the hierarchy, to move the limb from point A to point B in 3D workspace. With FK when the parent joint is moved within its hierarchy, all the children follow the parents rotation hence, the parent child relation is always maintained and so its termed as forward motion or kinematics. (**Fig. 1**) demonstrates the use of FK system to move the arm joints towards the glass by sequentially rotating each joint at appropriate angle towards its goal. For an animator this is a tedious task as the angle of each joint has to be tweaked to ensure the accurate placement of the arm.

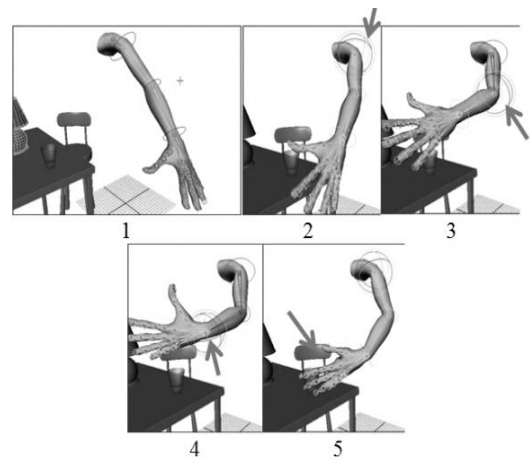

**Fig. 1: FK joint rotation order towards the glass on the table**

++ Corresponding Author: Zeeshan Bhatti, zeeshan.bhatti@live.iium.edu.my, +60-146222549

## 1.2 Inverse Kinematics (IK)

In IK system the animator simply specifies the final desired position of the limb using an end-effector and all the joints rotation are calculated automatically to place them at desired location. With IK move the last child in the hierarchy and all its parent joints will rotate, in Inverse motion or kinematics. The IK system in contrast to FK, uses a much direct approach in this case as shown in **(Fig. 2)**. The IK end-effector is directly translated to its desired place, in Figure 2 at the glass and all the joint rotations are computed and adjusted automatically according to the placement of the IK end-effector.

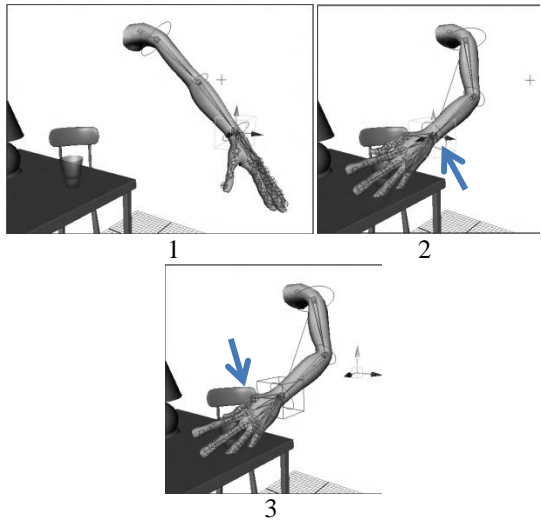

**Fig.2: IK Translation of Arm towards the glass on the table**.

It may seem that IK system is appropriate for moving arm or leg limbs but in situations where the character is falling, or just swinging its limbs then the IK proves to yield incorrect tangent and arcs for realistic motion, in such cases FK system is required. Thus in a 3D production environment both IK and FK approaches are used immensely and are equally important.

## 1.3 The problem with current systems.

The problem with using both these systems is that they cannot be used simultaneously or concurrently as IK systems needs explicit control over the joint structure, therefore once the joint have IK attached to them, they no longer are able to rotate with plain FK system and similarly, when FK joint rotations are required, then the joint cannot have IK.

One way to solve this problem is to use a 3 joint hierarchy structure of each limb. The first joint chain is controlled by FK system, the second joint chain is controlled by IK system and the third joint chain is attached to the mesh for deformation. Both IK and FK joint chains are constrained to the third joint chain and a switch is used to specify which control system the animator wants to use (Allen, *et al.,* 2008). The problem with this methodology is that being a complex joint structure, the switching between FK or IK system creates a popping or jumping effect. For example, when the animator translates the arm using IK, the FK is disabled and the arm is moved to its desired place, but when the animator switches back to FK, the third joint jumps back the original position of FK joint chain. Then when the animator again switches to IK, the FK joint becomes inactive and the third joint chain jumps to its previous IK position. Thus this approach is inappropriate and not feasible.

## 1.4 The proposed system

In this research work the system of single joint chain with both IK and FK system is implemented on same hierarchy. The user selects the system to be used and internally the switch is made preserving the initial and final location for both IK and FK without creating the popping or jumping effect, producing a seamless transition between the two systems. The **(Fig. 3)** shows the overview of the approach. We use Jacobian to determine the joint rotations and location between the IK and FK switch automatically and this way moving the corresponding kinematic joint system to the desired place automatically eliminating the jumping/popping effect and reducing the complexity of the system.

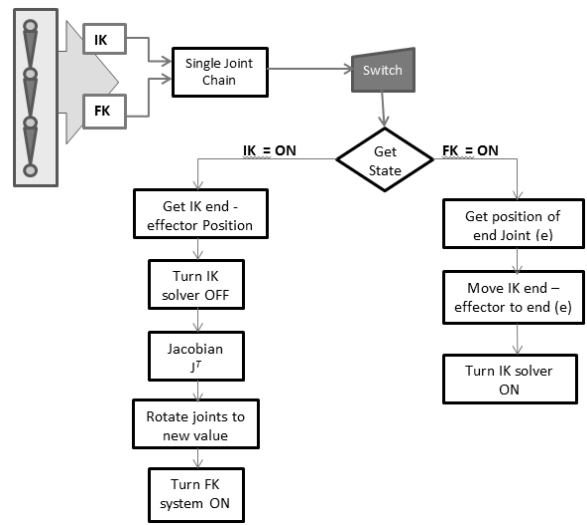

**Fig. 3: Overview of the IK / FK matching system**

## 2. MATERIAL AND METHOD

Our technique uses a single joint chain driven by both FK and IK system with a switch parameter to determine the currently active system as shown in Fig. 3, both FK and IK are built on the same joint chain. When FK system is active the IK is disabled automatically and when the IK system is used then the FK system becomes inactive by the system. When the system is switched from FK to IK, we simply determine the position of the end of the joint or the last joint in the system, and move the IK end-effector to the same coordinates in 3D space. Then the IK solver is enabled and the joint remain in the same position without popping effect. When the system is switched from the IK to FK, now the real problem occurs, as the joints were controlled by the position of end-effector as illustrated in **(Fig. 4)**. When the end-effector becomes inactive, the joint will go back to their original FK position. To solve this we used Jacobian transpose mechanism to determine the joint positions and then rotate each joint automatically to their new positions relevant to the IK effector, thus creating a seamless match and solving the problem.

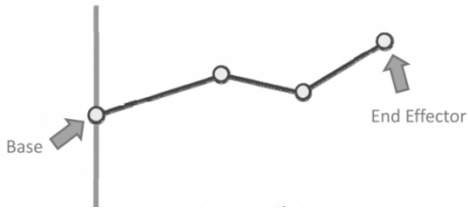

**Fig. 4: IK chain with base and end-effector**

### 3. COMPUTING FORWARD KINEMATICS

In FK the joints are always rotated towards their desired goal manually by the animator. The traversal of the joint chain follows a depth-first pattern from the root to leaf node. The process of computing FK motion in world space coordination is based on joint Degree Of Freedom (DOF) values as shown in **(Fig. 5)**.

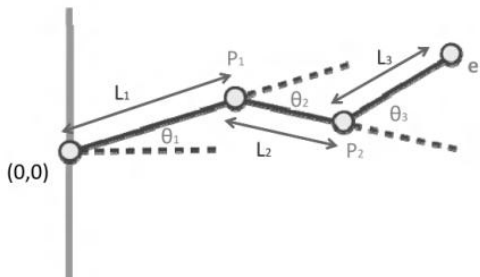

**Fig. 5: A joint structure using FK system for motion in world space.**

We begin calculation by first considering the DOF values for the joint $\theta = [\theta_1, \theta_2, \ldots, \theta_M]$ and the end joint in world space is described as $e = [e_1, e_2, \ldots, e_N]$ the motion of the FK end effector is computed indirectly as the accumulation of all transformations that lead to that end effector joint as (Watt, *et al.*, 1992) given in equation (1).

$$e = f(\theta) \qquad (1)$$

### 4. COMPUTING INVERSE KINEMATICS

To determine the position of upper joints of the hierarchy when the user switches the IK system to FK mode with end-effector at a specific goal position, we use the *Jacobian* which involves forming the matrix of partial derivatives (Parent, R., 2008). The concept of inverse kinematics is to compute the vector of joint DOFs that will cause the end effector to reach some desired goal state as shown in **(Fig. 6)**.

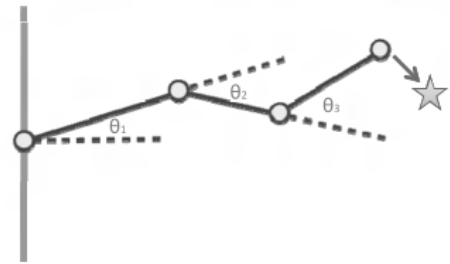

**Fig. 6: IK end-effector moved towards a goal.**

As we already have $e = [e_1, e_2, \ldots, e_N]$ from the FK system, and we want $\theta = [\theta_1, \theta_2, \ldots, \theta_M]$, therefore to determine the joint angle inversely by using equation (2).

$$\theta = f^{-1}(e) \qquad (2)$$

Inverse kinematics gives us the translation of the joint chain with linear approximation as compared to FK, in which the joint rotations yield a curved arc trajectory as shown in **(Fig.7)**. The FK position of the joint chain can be achieved through the IK end effector by computing the matrix of partial derivatives of entire system. The Jacobian defines how the end effector '**e**' changes relative to instantaneous changes in the system.

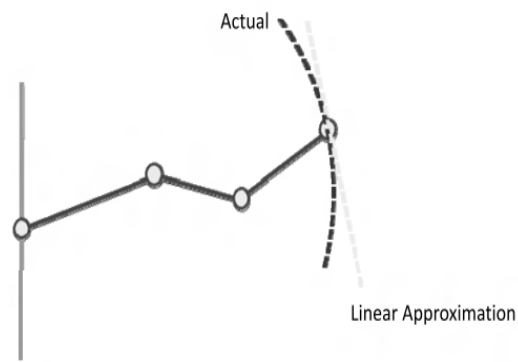

**Fig.7: FK and IK trajectory**

$$J = \frac{de}{d\theta} \qquad de = J\, d\theta \qquad (3)$$

$$e = [e_x\ e_y\ e_z]^T$$
$$\theta = [\theta_1\ \theta_2\ \ldots\ \theta_M]^T$$

The Jacobian Matrix thus is obtained as

$$J = \begin{bmatrix} \frac{\partial e_x}{\partial \theta_1} & \frac{\partial e_x}{\partial \theta_2} & \cdots & \frac{\partial e_x}{\partial \theta_M} \\ \frac{\partial e_y}{\partial \theta_1} & \frac{\partial e_y}{\partial \theta_2} & \cdots & \frac{\partial e_y}{\partial \theta_M} \\ \frac{\partial e_z}{\partial \theta_1} & \frac{\partial e_z}{\partial \theta_2} & \cdots & \frac{\partial e_z}{\partial \theta_M} \end{bmatrix} \qquad (4)$$

Recall that **θ = f⁻¹(e)** and **de = J dθ**, thus we obtained **dθ = J⁻¹ de.**

### 4.1 Computing Jacobian numerically
Let us examine one column of the Jacobian Matrix

$$\frac{\partial e}{\partial \theta_1} = \begin{bmatrix} \frac{\partial e_x}{\partial \theta_1} & \frac{\partial e_y}{\partial \theta_1} & \frac{\partial e_z}{\partial \theta_1} \end{bmatrix}^T$$

We can add a small Δθ to θ$_i$, then we can calculate how the end effector moves: Δe = e' – e. Now we have equation (5) which can be used to fill the Jacobian Matrix.

$$\frac{\partial e}{\partial \theta_1} \approx \frac{\Delta e}{\Delta \theta} = \begin{bmatrix} \frac{\Delta e_x}{\Delta \theta} & \frac{\Delta e_y}{\Delta \theta} & \frac{\Delta e_z}{\partial \theta} \end{bmatrix}^T \qquad (5)$$

### 4.2 Inverting the Jacobian Pseudoinverse
We then used the pseudo inverse to find a matrix that effectively inverts a non-square matrix by further solving the equation 3 as we already have computed Jacobian

de = J · dθ
J$^T$ · de = J$^T$J · dθ
(J$^T$J)$^{-1}$ J$^T$ · de = (J$^T$J)$^{-1}$ (J$^T$J)· dθ
(J$^T$J)$^{-1}$J$^T$ · de = dθ
J$^+$ · de = dΔθ
J$^+$ = (J$^T$J)$^{-1}$ J$^T$        (6)

Equation (6) now represents the pseudoinverse of *J* and it maps the desired motion of the end effector to the required changes of the joint angles.

### 4.3 Inverting the Jacobian— Jacobian Transpose
Another technique is just to use the transpose of the Jacobian matrix. As the Jacobian is already an approximation to f() it is much faster, but the quality of the resultant is found to be inaccurate and thus for our solution we preferred quality over performance, therefore pseudo inverse method was implemented to produce higher quality results.

### 4.4 Solving IK—Incremental Changes
Forward Kinematics is a nonlinear type of motion which implies that the Jacobian can only be used as an approximation that is valid near the current configuration, so the process of computing the Jacobian must be repeated while taking small iterative steps towards the target goal until we reach the desired location. The final algorithm for the said process is given in **(Fig 8).**

> Algorithm: Solving IK—Algorithm of the Jacobian Method
> 1. **while** (e is too far from g) {
> 2.     **compute** the Jacobian matrix J
> 3.     **c**ompute the pseudoinverse of the Jacobian
>        matrix **J⁺**
> 4.     **compute** the change in joint DOFs:
>        **Δθ = J⁺ · Δe**
> 5.     **apply** the changes to DOFs, move a small
>        step of **αΔθ: θ = θ + αΔθ**
> 6. }

**Fig 8: Algorithm of Solving IK using Jacobian Method**

### 5. RESULTS
The implementation of our system provides an efficinet and practically feasible system of switching between an inverse kinematics and forward kinematics based joint rig of any 3D virtual charcters. The initial biped control rig was created using the procedural mechanism descriped in our previous work (Bhatti, *et al.,* 2012). As our system uses single joint chain the complexity of creating the rig is reduced and consequently the time tradeoff in computing the motion of the rig is also automaticaly decreased, hence improving the overall performace of the virtual charcter during animation process. According to our system the user initally can start working with the character rig in either IK or FK mode. Then all that is required is to change a single attribute in the system 'Enable IK' to either 'On' for IK mode or set tthis variable to 'Off' for FK mode of motion as shown in **(Fig. 9(a-g).**

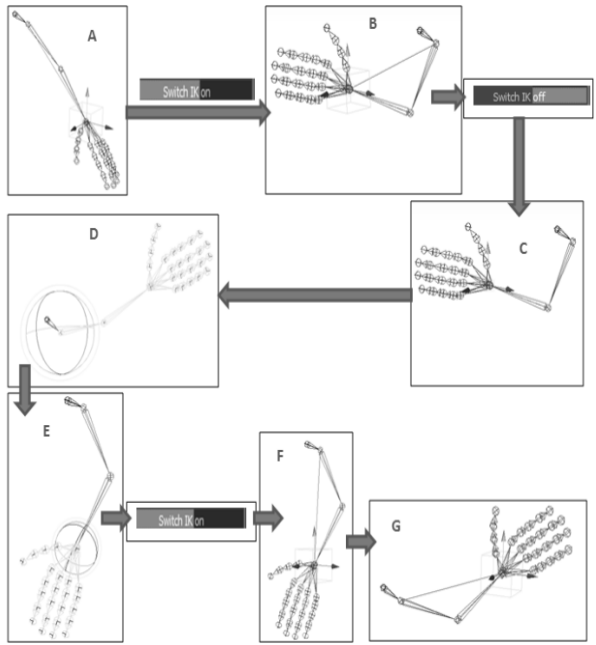

**Fig. 9:** The steps of switching arm rig between inverse kinematics and forward kinematics are shown from (A to G)

In Fig. 9-A the arm rig is in standard rest pose, then the 'switch IK' attribute is turned on to move the arm using Inverse kinematics as shown in Figure 9-B. then the 'Switch IK' attribute is turned off and the rig controls change to forwad kinamtics without altering the position of the arm joint as shown in Figure 9-C. From here the animator can easily continue to animate the arm using forwad kinematics as shown in figure 9-D and 9-E where joint ae rotated through forwad kinamatics. Finnally the 'Switch IK' attribute is again turned on to transfere the rig control back to inverse kinematics and thus by doing so the rig again remains exactly at the same place without altering its location, and the IK controls appear and the animator can safely animate the rig now using the IK movement. This process can be repeated and the animator can easily animate and chooses between various types of kinematics movements based on the reqirements without worrying to alter the rig.

**6.**  **CONCLUSION**

In this paper we attempted to solve the real world problem of IK to FK matching of joint chain for a 3D animated character without creating the popping or jumping effect. We used the mechanism of finding the Jacobian matrix $J$ and then computing its pseudoinverse $J^+$ to map the orientation values of IK joint chain to FK joint and similarly from FK to IK joints. The procedure allowed us to obtain the automation desired to do the seamless matching between the two types. As mathematical procedure is already proved its application in this situation gave us the solution and proved to be efficient also. The working algorithm was developed to implement the said technique in the animation of 3D articulated characters with both IK and FK system simultaneously. The work in future can be extended to use a full-body inverse and forward kinematics system with easy switching mechanism. The complexity of the algorithm can be further explored and minimized. The switching mechanism can also be implemented in quadruped joint structure with advance controls.

**REFERENCES:**

Allen, E., K. L Murdock, "Body Language: Advance 3D Character Rigging", Chapter 6, Cybex, Wiley Publishing, 2008, ISBN: 978-0-470-17387-9.

Apriantono T., H. Nunome, Y. Ikegami, S. Sano, (2006) "The effect of muscle fatigue on instep kicking kinetics and kinematics in association football" Journal of Sports Sciences Vol. **(24):** Issue. 9, 2006.

Bhatti, Z., and A. Shah, (2012) "Widget based Automated Rigging of Bipedal Character with Custom Manipulators", Proceedings of 11th ACM SIGGRAPH International Conference on Virtual-Reality Continuum and its Applications in Industry (VRCAI 2012), pages 337-340. ACM Press. ACM, New York, USA,. DOI=10.1145/2407516.2407593 http://doi.acm.org/10.1145/2407516.2407593 Dec 2-4, 2012.

Bruce A. Mac. Williams, M. Cowley, Diane E. Nicholson, (2003) Foot kinematics and kinetics during adolescent gait, Gait & Posture, Vol. **(17):** Issue 3, June 2003, Pages 214-224, ISSN 0966-6362, 10.1016/S0966-6362(02)00103-0. (http://www.sciencedirect.com/science/article/pii/S0966636202001030).

Grochow, K., Martin, S. L. Hertzmann, A. and Popović, Z. (2004) Style-based inverse kinematics. In ACM SIGGRAPH 2004 Papers (SIGGRAPH '04), Joe Marks (Ed.). ACM, New York, NY, USA, 522-531.DOI=10.1145/1186562.1015755 http://doi.acm.org/ 10.1145/1186562.1015755.

Jonathan D. (2008) Chappell, Orr Limpisvasti, Effect of a Neuromuscular Training Program on the Kinetics and Kinematics of Jumping Tasks, American Journal of Sports Medicine, vol. **(36):** no 6, 1081-1086; 2008,

Kuffner, J., K. Nishiwaki, S. Kagami, M. Inaba, and H. Inoue, "Motion planning for humanoid robots". In Robotics Research 365-374. Springer Berlin Heidelberg (2005).

Kulpa, R., and F. Multon, "Fast inverse kinematics and kinetics solver for human-like figures" Proceedings of 2005 5th IEEE-RAS International Conference on Humanoid Robot, Tsukuba, Japan, 0-7803-9320-1/05 IEEE-Xplore.

Li, L. Y., W. A. Gruver, and Q. X. Zhang, (2001) "Kinematic control of redundant robots and the motion optimizability measure", IEEE Transaction on Systems, Man, and Cyberrnatrics. Part B, Cybermatrics, vol. **(31):** no. 1, 155 -160,.

Parent, R., (2008) "Computer Animation: Algorithms and Techniques", Chapter 5, 187-215, second Edition, Elsevier-Morgan Kaufmann Publishers, ISBN: 978-0-12-532000-9,

Robin M. Queen, T. Michael Gross, Hsin-Yi Liu, (2006) Repeatability of lower extremity kinetics and kinematics for standardized and self-selected running speeds, Gait and Posture, Vol. **(23)** Issue 282-287, ISSN 0966-6362, 10.1016/j.gaitpost.2005.03.007. (http://www.sciencedirect.com/science/article/pii/S096 663620500069X)

Roepstorff, L., C. Johnston, and S. Drevemo, (1999) The effect of shoeing on kinetics and kinematics during the stance phase. Equine Veterinary Jour. 31: 279–285. doi: 10.1111/j.2042-3306.1999.tb05235.x

Rose III, C. F., P. P. J. Sloan & M. F. Cohen, (2001) September). Artist-Directed Inverse-Kinematics Using Radial Basis Function Interpolation. In Computer Graphics Forum Vol. **(20):** No. 3, 239-250. Blackwell Publishers Ltd.

Stirling L, K Willcox P. Ferguson D. Newman (2009) Kinetics and kinematics for translational motions in microgravity during parabolic flight. Aviation, Space, Environmental Medicine, Vol. **(80):** 6, J. 522-531.(10)

Tejomurtula, S., and S. Kak, (1999) Inverse kinematics in robotics using neural networks, Information Sciences, Vol. 116, Issues 2–4, 147-164, ISSN 0020-0255, 10.1016/S0020-0255(98)10098-1. (http://www.sciencedirect.com/science/article/pii/S002 0025598100981)

Watt, A., M. Watt, (1992) "Advance Animation and Rendering Techniques: Theory and Practice", Chapter 16, ACM Press, Addison-Wesley, ISBN: 0-201-54412-1

Xia, Y.S., G. Feng; J. Wang, (2005) "A primal-dual neural network for online resolving constrained kinematic redundancy in robot motion control," Systems, Man, and Cybernetics, Part B: Cybernetics, IEEE Transactions on , vol. 35, no.1, . 54-64, doi: 10.1109/TSMCB.2004.839913nURL: http://ieeexplore.ieee.org/stamp/stamp.jsp?tp=&arnum ber=1386426&isnumber=30179.